# Scalar Differential Equations of Thin Film Bulk Acoustic Wave Resonators with Surface Acoustic Impedance for Sensor Application

Huijing He[1], Jiashi Yang[2], and John A. Kosinski[3]*

[1]Department of Earth and Planetary Sciences, University of California, Santa Cruz, Santa Cruz, CA 95064, USA (email: hhe26@ucsc.edu)

[2]Department of Mechanical and Materials Engineering, University of Nebraska-Lincoln, Lincoln, NE 68588-0526, USA (e-mail: jyang1@unl.edu)

[3]Advanced Technology Group, MacAulay-Brown, Inc., Dayton, OH 45430 (e-mail: j.a.kosinski@ieee.org)

*Fellow, IEEE

*Abstract* – We generalize the two-dimensional scalar differential equations for the thickness-extensional operating modes of thin film bulk acoustic wave resonators (FBARs) to include the effects of surface impedance so that the equations can be used to model sensors based on FBARs. A mass sensor is analyzed as an example to show the effectiveness of the equations obtained. The mass layer inducer frequency shifts are calculated. It is also found that the vibration may be confined under the driving electrode or the mass layer. These are fundamental to mass sensor design.

## I. INTRODUCTION

Acoustic wave resonators are key components of oscillators as frequency standards for time keeping and signal processing used in many electronic equipment. Quartz crystal resonators (QCRs) [1] have been used for quite a few decades. Relatively recently, thin film bulk acoustic wave resonators (FBARs) [2, 3] made from ZnO or AlN on a silicon substrate have been developed for high frequency applications. Most QCRs and FBARs operate with thickness-shear or thickness-extensional modes [4] of crystal plates. These modes have spatial variations along the plate thickness only, and do not have any in-plane variations. Strictly speaking, thickness modes exist in unbounded plates without edge effects only. However, real devices are finite plates with boundaries. In finite plates the operating thickness modes have slow in-plane variations and the related edge effects. In-plane mode variations are also crucial in what is called energy trapping [1] which crucial to device mounting. With energy trapping, the vibration is confined near the center of the plate so that the device can be mounted at the edges without affecting its performance. To model in-plane mode variations for minimizing edge effects and designing energy trapping, researchers have developed two-dimensional scalar differential equations for the widely used AT-cut [5-8] and SC-cut QCRs [9-11]. The scalar equations are simple and accurate, and have been widely used to study QCRs, e.g., in [12-26]. Similar scalar differential equations have also been established for FBARs [27], which have been used effectively in FBAR theoretical analysis [28-33].

QCMs and FBARs have also been widely used to make mass, fluid, gas and biosensors. In these applications, resonators interact with additional mass layers, liquids or gasses on their surfaces. The scalar equations for QCRs and FBARs in [5-11, 27] cannot be used for sensor analysis and need to be generalized. In [34, 35], the scalar equations for AT-cut and SC-cut QCRs were generalized for QCR sensor applications through the consideration of surface acoustic impedance produced by the surface mass layers or fluids. In this paper we generalize the scalar equations for FBARs in [27] to include surface acoustic impedance for FBAR sensor application.

## II. GOVERNING EQUATIONS

Figure 1 shows the structure of a class of FBAR resonators. $X_1$ and $X_2$ are the in-plane coordinates. $X_3$ is the thickness coordinate. The FBAR consists of a piezoelectric film of ZnO or AlN and a silicon substrate, with a ground and a driving electrode. The driving electrode covers the central part of the resonator only. The top surface of the FBAR, whether there is a driving electrode or not, may be interacting with a mass layer or a fluid.

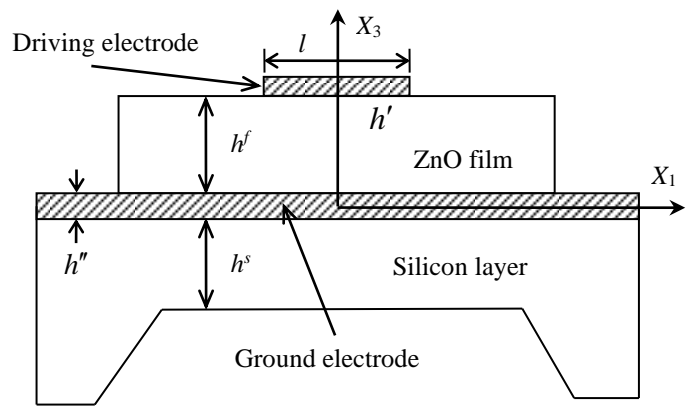

Fig.1. Schematic diagram of a typical ZnO-Si FBAR sensor.

The behavior of the piezoelectric film is governed by the three-dimensional theory of piezoelectricity [4] which is summarized briefly bellow. In the Cartesian tensor notation, the equation of motion and the charge equation of electrostatics are

$$T_{ji,j} = \rho \ddot{u}_i, \quad (1)$$
$$D_{i,i} = 0,$$

where **T** is the stress tensor, ρ the mass density, **u** the mechanical displacement, and **D** the electric displacement. (1) is accompanied by the following constitutive relations

$$T_{ij} = c_{ijkl} S_{kl} - e_{kij} E_k, \quad (2)$$
$$D_i = e_{ikl} S_{kl} + \varepsilon_{ik} E_k,$$

where **S** is the strain tensor and **E** the electric field. $c_{ijkl}$, $e_{ijk}$ and $\varepsilon_{ij}$ are the elastic, piezoelectric and dielectric constants. S and E are related to **u** and the electric potential $\varphi$ through

$$S_{kl} = (u_{l,k} + u_{k,l})/2, \quad (3)$$
$$E_i = -\varphi_{,i}.$$

The substrate is nonpiezoelectric. It is governed by equations similar to (1)-(3) but the piezoelectric constants are zero. The electrodes are assumed to be much thinner than the piezoelectric film and the substrate. Their inertia will be considered but the stiffness is neglected [27]. Thus the effects of the electrodes appear in the boundary and continuity (jump) conditions below.

At the top of the FBAR, its interaction with a mass layer or a fluid is represented by an acoustic impedance matrix $Z_{ij}$. In the region without a driving electrode, the boundary conditions are

$$T_{3j} = -Z_{ji} \dot{u}_i, \quad D_3 = 0, \quad \text{at } x_3 = h^f. \quad (4)$$

In the region with a driving electrode, let the driving voltage be *V*, the boundary conditions are

$$\varphi = V \exp(i\omega t), \quad \text{at } x_3 = h^f, \quad (5)$$
$$T_{3j} = -\rho' h' \ddot{u}_j - Z_{ji} \dot{u}_i, \quad \text{at } x_3 = h^f,$$

where $\rho'$ and $h'$ are the density and thickness of the driving electrode. At the interface between the piezoelectric film and the substrate, the continuity conditions are

$$u_j^f = u_j^s, \quad \varphi = 0, \quad \text{at } x_3 = 0, \quad (6)$$
$$T_{3j}^f - T_{3j}^s = \rho'' h'' \ddot{u}_j, \quad \text{at } x_3 = 0,$$

where $\rho''$ and $h''$ are the density and thickness of the ground electrode. The superscripts "*f*" and "*s*" are for the film and the substrate, respectively. Finally, at the bottom of the FBAR, we have

$$T_{3j} = 0, \quad \text{at } x_3 = -h^s. \quad (7)$$

## III. SCALAR EQUATIONS

We consider time-harmonic motions and use the usual complex notation. All fields have the same time dependence with a common factor $\exp(i\omega t)$ which will be dropped when there is no confusion. The *n*th-order thickness-extensional displacement may be approximately represented by

$$u_3^n \cong f^n(x_1, x_2) g^n(x_3) \exp(i\omega t), \quad (8)$$

in which the slow in-plane field variation is described by $f(x_1, x_2)$, and the thickness dependence by $g(x_3)$. From (1)-(7), through very lengthy algebra [36], approximate dispersion relations can be obtained for long in-plane waves accurate up to the second order of the small in-plane wave numbers. Then second-order two-dimensional scalar differential equations can be constructed for $f^n$. They are slightly different depending on whether there is a driving electrode or not. When there is a driving electrode,

$$M_n\left(\frac{\partial^2 f^n}{\partial x_1^2} + \frac{\partial^2 f^n}{\partial x_2^2}\right) - \bar{c}_{33}^f(\hat{\eta}_{fn})^2 f^n - \rho^f \ddot{f}^n = 0, \tag{9}$$

where the thickness wavenumber is given by

$$\hat{\eta}_{fn} = \eta_{fn}^0 (1 + \frac{P^0}{G^0}), \tag{10}$$

and

$$P^0 = \frac{k^2}{(\eta_{fn}^0 h^f)^2}\left[\frac{2}{\cos(\eta_{fn}^0 h^f)} - 2 + c^r \mu \tan(\eta_{fn}^0 h^f)\tan(\mu\sigma\eta_{fn}^0 h^f)\right] - R''$$
$$-\left[R' - \frac{i\omega h^f Z_{33}}{\bar{c}_{33}^f (\eta_{fn}^0 h^f)^2}\right][1 - c^r \mu \tan(\eta_{fn}^0 h^f)\tan(\mu\sigma\eta_{fn}^0 h^f)], \tag{11}$$

$$G^0 = \sec^2(\eta_{fn}^0 h^f) + c^r \mu^2 \sigma \sec^2(\mu\sigma\eta_{fn}^0 h^f). \tag{12}$$

In Eqs. (11) and (12), the thickness wavenumber $\eta_{fn}^0$ is introduced to denote the $n$-th harmonic vibration of a reference structure without electrodes and surface impedance. The detailed expressions of other variables are given in Appendix A.

When there is not a driving electrode,

$$M_n\left(\frac{\partial^2 f^n}{\partial x_1^2} + \frac{\partial^2 f^n}{\partial x_2^2}\right) - \bar{c}_{33}^f(\bar{\eta}_{fn})^2 f^n - \rho^f \ddot{f}^n = 0, \tag{13}$$

where the thickness wavenumber is

$$\bar{\eta}_{fn} = \eta_{fn}^0\left[1 - \frac{R'' - \frac{i\omega Z_{33}}{\bar{c}_{33}^f (\eta_{fn}^0)^2 h^f}\sec^2(\eta_{fn}^0 h^f)}{c^r \mu^2 \sigma \sec^2(\mu\sigma\eta_{fn}^0 h^f) + \sec^2(\eta_{fn}^0 h^f)}\right], \tag{14}$$

For an electroded plate with driving voltage $V$ applied on its top electrode, the scalar equation takes the form

$$M_n\left(\frac{\partial^2 f^n}{\partial x_1^2} + \frac{\partial^2 f^n}{\partial x_2^2}\right) - \bar{c}_{33}^f(\hat{\eta}_{fn})^2 f^n - \rho^f \ddot{f}^n = \rho^f \omega_n^2 \frac{e_{33}^f}{c_{33}^f}\frac{A_1}{h^f(A_2+A_3)}V\exp(i\omega t), \tag{15}$$

where the parameters defining the driving term on the right hand side are

$$A_1 = \frac{\rho^f[1-\cos(\eta_{fn}^0 h^f)]}{(\eta_{fn}^0)^2 \sin(\eta_{fn}^0 h)}, \quad A_2 = \rho^f h^f \frac{\sin(2\eta_{fn} h^f) + 2\eta_{fn} h^f}{4\eta_{fn}\sin^2(\eta_{fn} h^f)},$$

$$A_3 = \rho^s h^s \left(\frac{\bar{c}_{33}^f \eta_{fn}^0}{c_{33}^s \eta_{sn}^0}\right)^2 \frac{2\eta_{sn}^0 h^s + \sin(2\eta_{sn}^0 h^s)}{4\eta_{sn}^0 \sin^2(\eta_{sn}^0 h^s)}. \tag{16}$$

In Eq. (15) we have omitted the small quantity of first-order in the mass of the electrodes, the electromechanical coupling factor and the surface impedance, which was denoted by $\delta$ in Eq. (4.4.43) in [36]. When the electromechanical coupling is strong, or the surface impedance is large, this term needs to be taken into account. When the surface impedance terms are set to zero, (9) and (13) reduce to the corresponding equations in [27]. The coefficients of the equations in (9) and (13) need to be calculated through a series of intermediate steps which are given in Appendix A.

## IV. ANALYSIS OF AN FBAR MASS SENSOR

As an example for the application of (9) and (13), we analyze the FBAR used as a mass sensor as shown in Fig. 2. We want to obtain free vibration ($V=0$) frequencies and modes.

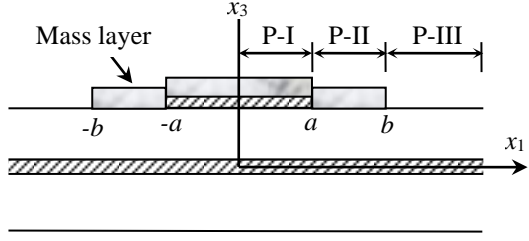

Fig. 2. An FBAR with a mass layer as a mass sensor.

The normal impedance $Z_{33}$ due to the surface mass layer can be determined in a manner similar to the tangential impedance in [34, 35]. The equation of motion in the $x_3$ direction of a differential element of the surface mass layer with a unit area is

$$-T_{33} = \rho^l h^l \ddot{u}_3 = \rho^l h^l i\omega \dot{u}_3, \tag{17}$$

where $\rho^l$ and $h^l$ are the mass layer density and thickness, respectively. Comparing (11) with the definition of $Z_{33}$, i.e.,

$$-T_{33} = Z_{33}\dot{u}_3, \tag{18}$$

we identify

$$Z_{33} = \begin{cases} i\omega \rho^l h^l, & |x_1|<b, \\ 0, & |x_1|>b. \end{cases} \tag{19}$$

The problem is symmetric about the $x_3$ axis. We work on the right half only and divide it into 3 regions. (9) is valid in [0, $a$], denoted by P-I, and (13) is valid in [$a$, $b$] and [$b$, ∞], denoted by P-II and P-III, respectively. The fields are bounded at the origin. We are interested in trapped modes that decay to zero at infinity. There are two possibilities. When the mass layer is very thin, the modes are trapped by the driving electrode and

$$f^n = \begin{cases} A\cos\xi_1 x_1, & 0<x_1<a, \\ B_1\exp(-\xi_2 x_1) + B_2\exp(\xi_2 x_1), & a<x_1<b, \\ C\exp(-\xi_3 x_1), & b<x_1<\infty. \end{cases} \tag{20}$$

where

$$\xi_1 = \sqrt{\frac{\rho^f \omega^2 - \bar{c}_{33}^f (\hat{\eta}_{fn})^2}{M_n}}, \tag{21a}$$

$$\xi_2 = \sqrt{\frac{\bar{c}_{33}^f (\bar{\eta}_{fn})^2 - \rho^f \omega^2}{M_n}} \quad \text{and } Z_{33} \neq 0, \tag{21b}$$

$$\xi_3 = \sqrt{\frac{\bar{c}_{33}^f (\bar{\eta}_{fn})^2 - \rho^f \omega^2}{M_n}}, \quad \text{and } Z_{33} = 0. \tag{21c}$$

The frequency equation is obtained by applying the continuity condition of both $f^n$ and its derivative at $x_3 = a$ and $b$, which is given by

$$\begin{bmatrix} \cos(\xi_1 a) & -\exp(-\xi_2 a) & -\exp(\xi_2 a) & 0 \\ -\xi_1 \sin(\xi_1 a) & \xi_2 \exp(-\xi_2 a) & -\xi_2 \exp(\xi_2 a) & 0 \\ 0 & \exp(-\xi_2 b) & \exp(\xi_2 b) & -\exp(-\xi_3 b) \\ 0 & -\xi_2 \exp(-\xi_2 b) & \xi_2 \exp(\xi_2 b) & \xi_3 \exp(-\xi_3 b) \end{bmatrix} = \mathbf{0}. \tag{22}$$

When the mass layer is not very thin, the modes are trapped by the mass layer and

$$f^n = \begin{cases} A\cos\xi_1 x_1, & 0 < x_1 < a, \\ B_1\cos\xi_2 x_1 + B_2\sin\xi_2 x_1, & a < x_1 < b, \\ C\exp(-\xi_3 x_1), & b < x_1 < \infty. \end{cases} \quad (23)$$

where $\xi_1$ and $\xi_3$ are given by Eqs. (21a) and (21c), and

$$\xi_2 = \sqrt{\frac{\rho^f \omega^2 - \overline{c}_{33}^f (\overline{\eta}_{fn})^2}{M_n}} \quad \text{and} \quad Z_{33} \neq 0. \quad (24)$$

Correspondingly, the frequency equation is given by

$$\begin{bmatrix} \cos(\xi_1 a) & -\cos(\xi_2 a) & -\sin(\xi_2 a) & 0 \\ -\xi_1 \sin(\xi_1 a) & \xi_2 \sin(\xi_2 a) & -\xi_2 \cos(\xi_2 a) & 0 \\ 0 & \cos(\xi_2 b) & \sin(\xi_2 b) & -\exp(-\xi_3 b) \\ 0 & -\xi_2 \sin(\xi_2 b) & \xi_2 \cos(\xi_2 b) & \xi_3 \exp(-\xi_3 b) \end{bmatrix} = \mathbf{0}. \quad (25)$$

## V. NUMERICAL RESULTS AND DISCUSSION

In this section we present several numerical examples to show the potential applications of the scalar differential equation in analysis and optimization of FBAR sensor by analyzing its frequency behavior and energy trapping properties. We consider an FBAR mass sensor consists of a ZnO thin film deposited on a Silicon substrate. The thickness of the ZnO film is 15 μm, and the thickness of the Si substrate is 5 μm. The ground electrode is made of 200 nm thick gold layer with areal mass ratio R"=0.0453, the surface driving electrode covering the center portion of the composite structure has areal mass ratio R'=0.0453. The mechanical and electrical properties of ZnO and Si are shown in Tab. 1. The length of the surface electrode is 2a= 0.76 mm, and the length of the deposited mass layer is 2b=1.5 mm. To simplify the discussion, we denote the central portion of the structure covered by the driving electrode and the mass loading by P-I, the middle part of the plate covered by the surface mass loading by P-II, and the unloaded segment by P-III, which are shown in Fig.2. Before investigating the energy trapping mode in an FBAR with finite length, we first study the dispersion characteristics of an infinite plate. For an infinite ZnO/Si plate without surface and ground electrodes, and in an unloaded state, the resonant frequency of its fundamental thickness-extensional mode is $\omega_0^{(1)} = 709.99677 \times 10^6$ rad/s. We use this frequency to normalize the frequency of electroded plate with or without surface mass loading, which is defined by $\Omega = \omega / \omega_0^{(1)}$. The definition of the in-plane wavenumbers of the plate with electrode and/or surface mass layer are given in Eqs. (21) and (24). For FBAR resonators used as a mass sensor, the thickness of the deposited surface layer is to be monitored by measuring the frequency change before and after its deposition. To quantify the effects of the surface impedance, which is an additional mass layer with unknown properties, we introduce a parameter called the areal mass density ratio, denoted by $R_Z$, $R_z = (\rho^l h^l)/(\rho^f h^f)$, which is related to the surface acoustic impedance by $Z_{33} = i\omega R_z \rho^f h^f$.

Table 1  Mechanical and electrical properties of the ZnO/Si FBAR

|  | Density: $\rho$ (kg/m$^3$) | $c_{33}$ (GPa) | $c_{13}$ (GPa) | $c_{44}$ (GPa) | $e_{33}$ (C/m$^2$) | $\varepsilon_{33}$ (F/m) |
|---|---|---|---|---|---|---|
| ZnO | 5675 | 211 | 105 | 43 | 1.57 | 96.51×10$^{-12}$ |
| Si | 2332 | 165.7 | 63.9 | 79.56 |  |  |

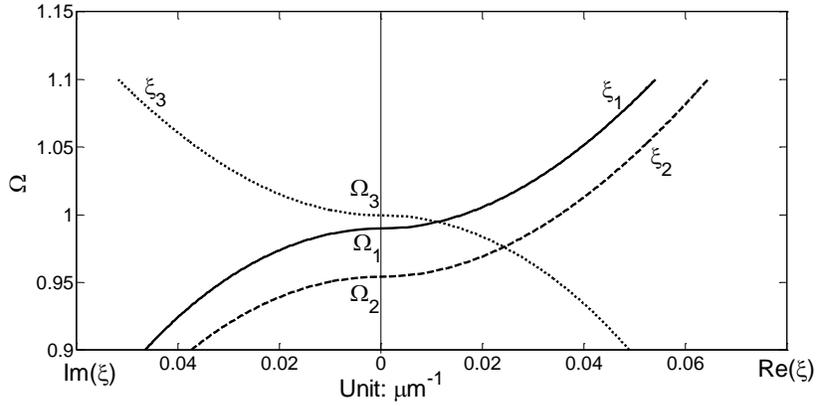

Figure 3 Dispersion curves near the fundamental mode of a partially electrode FBAR mass sensor

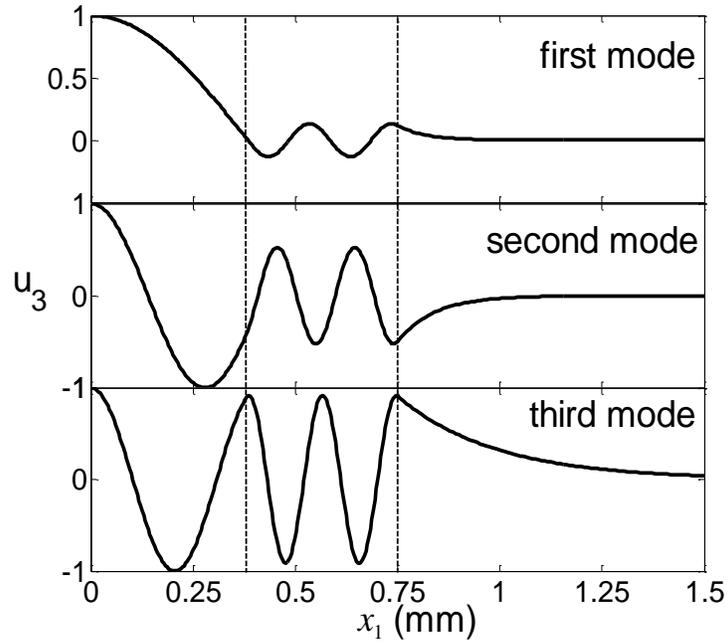

Figure 4 Mode shape of the first three modes near the fundamental thickness-extensional mode of an FBAR mass sensor

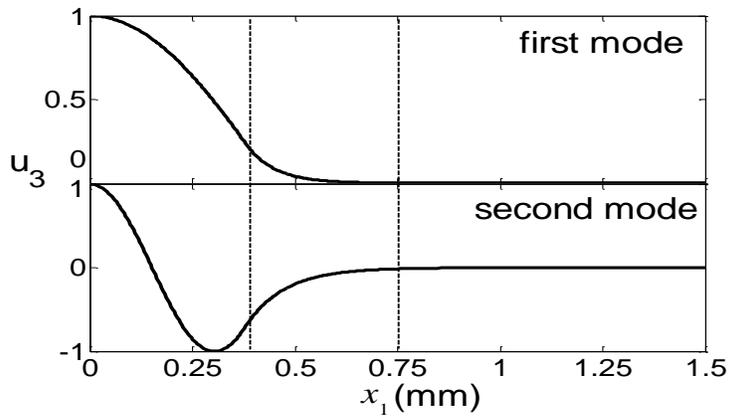

Figure 5 Mode shape of the first two modes near the fundamental thickness-extensional mode of a reference FBAR resonator

As a specific example, we choose $R_Z=0.05$ to show the frequency and mode characteristics of the FBAR sensor. The dispersion curves for an infinite plate with different surface loading conditions, for which the in-plane wavenumbers are given in Eqs. (20) and (24) are shown in Fig. 3. It is seen that energy trapping modes can only exist in the frequency range $\Omega_1 < \Omega < \Omega_3$. In addition, we can observe that propagating waves exist in region P-II. Solving for the resonant frequency from the frequency equation (25), we find three vibration modes, for which the eigenfrequencies are $\omega_1^{(1)} = 703.19985 \times 10^6$ rad/s, $\omega_2^{(1)} = 706.28879 \times 10^6$ rad/s and $\omega_3^{(1)} = 709.30584 \times 10^6$ rad/s. The corresponding vibration modes are shown in Fig. 4. The three regions with different surface conditions are marked by dashed lines. From the mode shapes we can observe that the first anharmonic mode has the best energy trapping since the most active region with the largest vibration amplitude is confined in the electroded part. The vibration amplitudes in the region P-II of the second and third anharmonic modes are comparable with the amplitude in P-I, although they are both energy trapping modes. To observe the changes in the frequency and mode shape caused by the mass loading, we plot the mode shape of unloaded plate in Fig. 5. We can see that the mass layer causes more energy to be distributed in the region far from the central electroded part. The eigenfrequencies for the unloaded FBAR can be solved from the frequency equation similarly. Only two energy trapping modes are found, of which the frequencies are $\tilde{\omega}_1^{(1)} = 703.94553 \times 10^6$ rad/s and $\tilde{\omega}_2^{(1)} = 706.62575 \times 10^6$ rad/s, respectively. The frequency shifts caused by the mass layer are $\Delta\omega_1^{(1)} = 0.74568 \times 10^6$ rad/s and $\Delta\omega_2^{(1)} = 0.33696 \times 10^6$ rad/s, respectively. We need to point out that the third mode shown in Fig. 4 is a new mode caused by the deposition of the mass layer, there is no corresponding mode for the unloaded plate. In order to study the sensitivity of the FBAR in more details, we further plot the dependence of the frequency of the first anharmonic mode $\omega_1^{(1)}$ on the mass loading $R_Z$, as shown in Fig. 6.

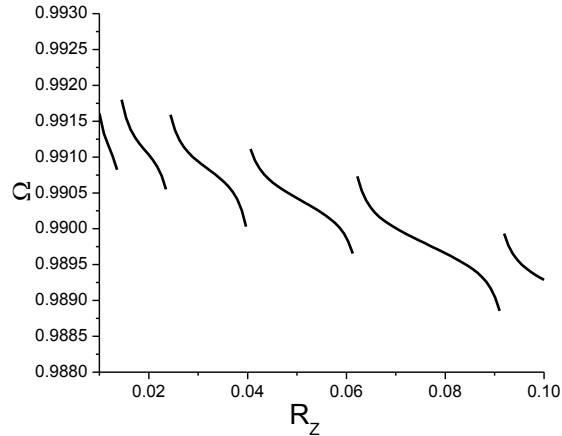

Figure 6 Frequency sensitivity curve of the first anharmonic near
the fundamental thickness extensional mode

From Fig. 6 we can observe that the sensitivity curve is broken into several segments, in each of them the frequency is a monotonically decrease function of mass loading. In practical applications we should adjust the parameters of the FBAR to avoid using the frequency band in which the frequency shift shows a jump.

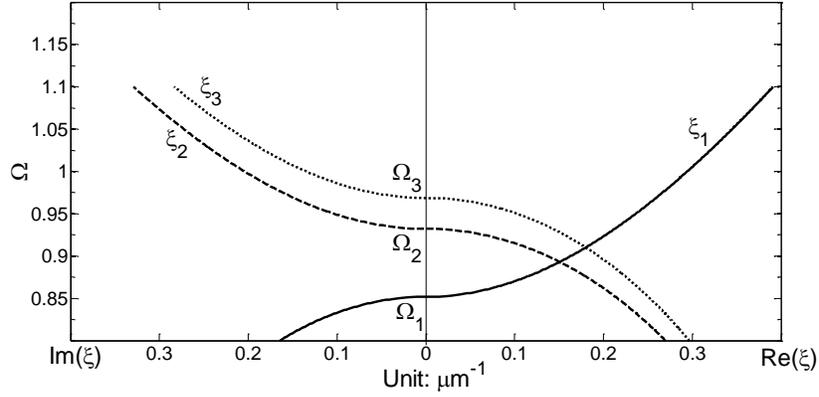

Figure 7 Dispersion curves near the second harmonic of a partially electrode FBAR mass sensor

There is no energy trapping mode near the first overtone of the thickness extensional mode. The dispersion curves of the second overtone of the thickness extensional mode of an infinite FBAR are shown in Fig. 7. For this case, the eigenfrequency of the second overtone of an infinite reference plate is $\omega_0^{(3)} = 1165.28844 \times 10^6$ rad/s. We normalize the frequency by $\Omega = \omega / \omega_0^{(3)}$. It is seen that energy trapping modes can exist in the region $\Omega_1 < \Omega < \Omega_3$. However, the wavenumber $\xi_1$ for modes in the frequency range $\Omega_2 < \Omega < \Omega_3$ is relatively large, indicating that these modes can hardly be excited due to the charge cancellation caused by the highly oscillating waves in the electroded part. Consequently, we only consider modes in the range $\Omega_1 < \Omega < \Omega_2$. In this case, attenuating waves exist in P-II. Solving the frequency equation (22) for the finite FBAR, it is found that there are 26 energy trapping modes in this frequency range. We plot the first two modes in Fig. 8. The eigenfrequencies of these modes are $\omega_1^{(3)} = 992.59521 \times 10^6$ rad/s and $\omega_2^{(3)} = 992.88400 \times 10^6$ rad/s, respectively. We also plot the mode shape of the unloaded plate in the figure. It is seen that the mode shapes are nearly indistinguishable from one another. Both modes show excellent energy trapping where nearly all the energy is confined in P-I. The resonant frequency of the unloaded plate can be solved from the corresponding frequency equation and we obtain $\tilde{\omega}_1^{(3)} = 1027.85337 \times 10^6$ rad/s and $\tilde{\omega}_2^{(3)} = 1028.15219 \times 10^6$ rad/s, which give frequency changes $\Delta\omega_1^{(3)} = 35.25816 \times 10^6$ rad/s and $\Delta\omega_2^{(3)} = 35.26819 \times 10^6$ rad/s, respectively. We can find that although the mode shapes are nearly the same, the frequency shifts are very large, significantly larger than those near the fundamental mode. This tells us that these modes are capable of providing much higher sensitivity while still perfectly retain the original mode shape.

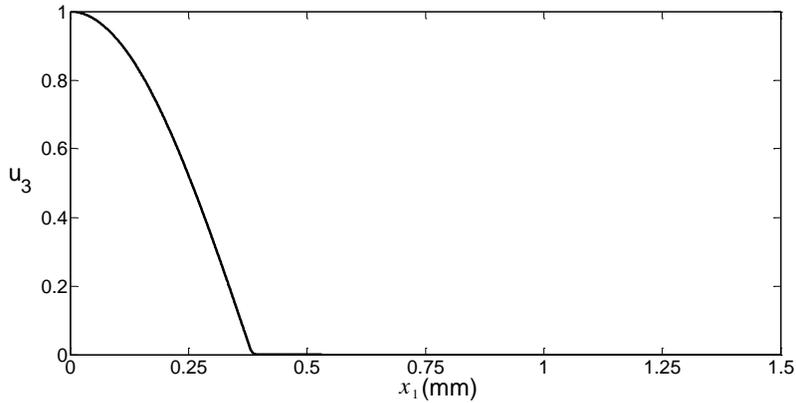

(a)

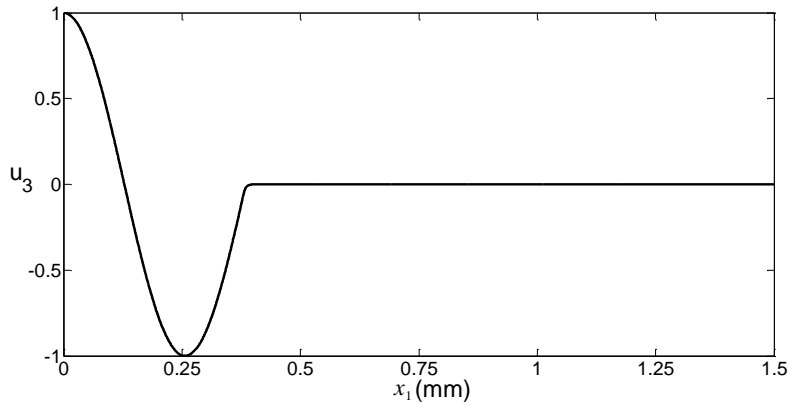

(b)

Figure 8 Vibration mode of the second harmonic of an FBAR mass sensor

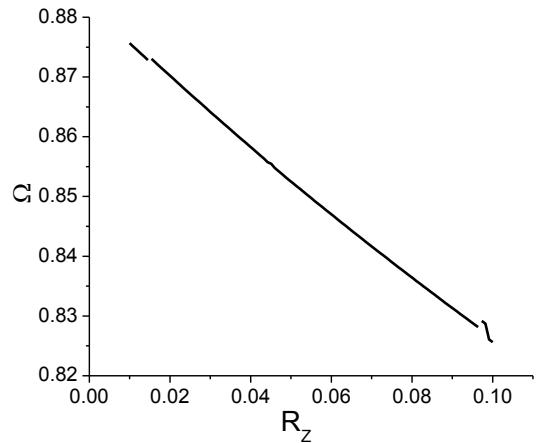

Figure 9 Frequency sensitivity of the first anharmonic near
the second overtone of the thickness extensional mode

Finally, we show the sensitivity curve of the first anharmonic mode near the second overtone in Fig. 9. It is seen that this mode has much larger frequency sensitivity compared to those near the fundamental mode. Moreover, the monotonically decreasing frequency band is much larger than that of the fundamental mode.

## VI. CONCLUSIONS

A system of scalar differential equations for the thin film bulk acoustic wave resonator sensors (FBARs) operating with the thickness extensional modes are developed. Numerical examples for an FBAR resonator used as a mass sensor show that the new equations are capable of predicting the frequency sensitivity and energy-trapping modes of such sensors, which are of great value for the design and optimization of FBAR sensors. Compared to the conventional method like finite element method, the scalar equations provide a computationally more efficient way to guide the analysis of this type of sensors.

## ACKNOWLEDGEMENT

This work was supported in part by the US Army Research Laboratory/US Army Research Office under agreement number W911NF-10-1-0293.## REFERENCES

APPENDIX

In this appendix we list the detailed expression of the parameters used in the scalar differential equation,

Piezoelectric stiffened stiffness: $\bar{c}_{33}^f = c_{33}^f + \frac{(e_{33}^f)^2}{\varepsilon_{33}^f}$

Ratio of the elastic stiffness of the silicon layer to that of the piezoelectric film: $c^r = \frac{c_{11}^s}{\bar{c}_{33}^f}$

Electromechanical coupling factor: $k^2 = \dfrac{(e_{33}^f)^2}{\bar{c}_{33}^f \varepsilon_{33}^f}$

Ratio of the areal mass density of the top electrode to that of the piezo-film: $R' = \dfrac{\rho' h'}{\rho^f h^f}$

Ratio of the areal mass density of the bottom electrode to that of the piezo-film: $R'' = \dfrac{\rho'' h''}{\rho^f h^f}$

Ratio of the longitudinal wave speed of the piezo-film to that of the silicon layer: $\mu = \sqrt{\dfrac{\rho^s \bar{c}_{33}^f}{\rho^f c_{11}^s}}$

Ratio of the thickness of the silicon layer to that of the piezo-film: $\sigma = \dfrac{h^s}{h^f}$

Thickness wavenumber of longitudinal waves in the piezo-film: $\eta_{f1}^0 \equiv \eta_f^0$, $(\eta_{f1}^0)^2 = \dfrac{\rho^f \omega_0^2}{\bar{c}_{33}^f}$

Thickness wavenumber of longitudinal waves in the silicon layer: $\eta_{s1}^0 \equiv \eta_s^0$, $(\eta_{s1}^0)^2 = \dfrac{\rho^s \omega_0^2}{c_{33}^s}$

Thickness wavenumber of shear waves in the piezo-film: $\eta_{f2}^0 = \kappa^f \eta_f^0$, $(\eta_{f2}^0)^2 = \dfrac{\rho^f \omega_0^2}{c_{44}^f}$

Thickness wavenumber of longitudinal waves in the silicon layer: $\eta_{s2}^0 = \kappa^s \eta_s^0$, $(\eta_{s2}^0)^2 = \dfrac{\rho^s \omega_0^2}{c_{44}^s}$, where $\kappa^f = \sqrt{\dfrac{\bar{c}_{33}^f}{c_{44}^f}}$, $\kappa^s = \sqrt{\dfrac{c_{33}^s}{c_{44}^s}}$.

The eigenfrequency $\omega_0$ and the corresponding thickness wavenumbers of the reference structure, i.e., neglecting the effects of all the electrodes and the surface impedances, are solved from the frequency equation

$$c^r \mu \tan(\mu \sigma \eta_f^0 h^f) + \tan(\eta_f^0 h^f) = 0 \ . \tag{A1}$$

This equation has an infinite number of solutions, which are called the fundamental, the first and higher harmonics starting from lower to higher values, denoted by $\eta_{fn}^0$, n=1, 2, 3, …

The coefficients of the partial derivative with respect to the in-plane coordinates $x_1$ and $x_2$ the scalar equations have the expression

$$M_n = g^f + 2\bar{c}_{33}^f \eta_{fn}^0 \dfrac{W}{h^f} \ , \tag{A2}$$

where

$W = \dfrac{P+VK}{R+V\mu\sigma}$,

$P = (l_{22}l_{44} - l_{24}l_{42})(l_{11}m_{33} - l_{13}m_{31} + m_{11}l_{33} - m_{13}l_{31}) + (l_{31}n_{43} - l_{33}n_{41})(n_{14}l_{22} - n_{12}l_{24}) +$
$\quad (n_{21}l_{33} - n_{23}l_{31})(n_{14}l_{42} - n_{12}l_{44}) + (n_{34}l_{22} - n_{32}l_{24})(n_{41}l_{13} - n_{43}l_{11}) + (n_{21}l_{13} - n_{23}l_{11})(n_{32}l_{44} - n_{34}l_{42})$,

$V = (l_{44}l_{22} - l_{42}l_{24})(p_{33}^s l_{11} - p_{31}l_{13})$,

$R = (l_{44}l_{22} - l_{42}l_{24})(p_{33}^f l_{11} + p_{11}l_{33} - l_{31}p_{13})$, $\quad K = \dfrac{\rho^s h^s}{2\eta_s^0 c_{33}^s}\left(\dfrac{g^s}{\rho^s} - \dfrac{g^f}{\rho^f}\right)$,

$g^f = r^f(c_{13}^f + c_{44}^f) + c_{44}^f$, $\quad g^s = r^s(c_{13}^s + c_{44}^s) + c_{44}^s$,

$r^f = \dfrac{c_{13}^f + c_{44}^f}{\bar{c}_{33}^f - c_{44}^f}$, $\quad r^s = \dfrac{c_{13}^s + c_{44}^s}{c_{33}^s - c_{44}^s}$,

$$l_{11} = -\bar{c}_{33}^f \eta_f^0 \sin(\eta_f^0 h^f), \ l_{13} = \bar{c}_{33}^f \eta_f^0 \cos(\eta_f^0 h^f), \ l_{22} = \eta_f^0 \cos(\eta_f^0 h^f), \ l_{24} = -\eta_{f2}^0 \sin(\eta_{f2}^0 h^f),$$

$$l_{31} = c_{33}^s \eta_s^0 \sin(\eta_s^0 h^s), \ l_{33} = \bar{c}_{33}^f \eta_f^0 \cos(\eta_s^0 h^s), \ l_{42} = \frac{c_{44}^f}{c_{44}^s} \eta_{f2}^0 \cos(\eta_{s2}^0 h^s), \ l_{44} = \eta_{s2}^0 \sin(\eta_{s2}^0 h^s),$$

$$m_{11} = \frac{c_{13}^f}{\eta_f^0} r^f \sin(\eta_f^0 h^f), \ m_{13} = -\frac{c_{13}^f}{\eta_f^0} r^f \cos(\eta_f^0 h^f), \ m_{22} = \frac{r^f}{\eta_{f2}^0} \cos(\eta_{f2}^0 h^f), \ m_{24} = -\frac{r^f}{\eta_{f2}^0} \sin(\eta_{f2}^0 h^f),$$

$$m_{31} = a_2 \sin(\eta_s^0 h^s) + a_1 \sin(\eta_{s2}^0 h^s), \ m_{33} = a_5 \cos(\eta_s^0 h^s) + a_6 \cos(\eta_{s2}^0 h^s),$$

$$m_{42} = b_2 \cos(\eta_s^0 h^s) + b_3 \cos(\eta_{s2}^0 h^s), \ m_{44} = b_5 \sin(\eta_{s2}^0 h^s) + b_6 \sin(\eta_s^0 h^s),$$

$$n_{12} = (\bar{c}_{33}^f r^f - c_{13}^f) \sin(\eta_{f2}^0 h^f), \ n_{14} = (\bar{c}_{33}^f r^f - c_{13}^f) \cos(\eta_{f2}^0 h^f),$$

$$n_{21} = -(r^f + 1) \cos(\eta_f^0 h^f), \ n_{23} = -(r^f + 1) \sin(\eta_f^0 h^f),$$

$$n_{32} = -a_3 \sin(\eta_s^0 h^s) + a_4 \sin(\eta_{s2}^0 h^s), \ n_{34} = (c_{33}^s r^s - c_{13}^s) \cos(\eta_{s2}^0 h^s) + a_7 \cos(\eta_s^0 h^s),$$

$$n_{41} = -(r^s + 1) \cos(\eta_s^0 h^s) + b_1 \cos(\eta_{s2}^0 h^s), \ n_{43} = \frac{\bar{c}_{33}^f}{c_{33}^s} \frac{\eta_f^0}{\eta_s^0} (r^s + 1) \sin(\eta_s^0 h^s) + b_4 \sin(\eta_{s2}^0 h^s),$$

$$p_{11} = -\bar{c}_{33}^f \left[ \eta_f^0 \cos(\eta_f^0 h^f) + \frac{1}{h^f} \sin(\eta_f^0 h^f) \right], \ p_{13} = \bar{c}_{33}^f \left[ \frac{1}{h^f} \cos(\eta_f^0 h^f) - \eta_f^0 \sin(\eta_f^0 h^f) \right],$$

$$p_{31} = c_{33}^s \left[ \eta_s^0 \cos(\eta_s^0 h^s) + \frac{1}{h^s} \sin(\eta_s^0 h^s) \right], \ p_{33}^f = \frac{\bar{c}_{33}^f}{h^f} \cos(\eta_s^0 h^s), \ p_{33}^s = -\bar{c}_{33}^f \eta_f^0 \sin(\eta_s^0 h^s),$$

$$a_1 = \frac{1}{c_{44}^s \eta_{s2}^0} [c_{44}^s (r^s + 1) - c_{44}^f (r^f + 1)](c_{13}^s - r^s c_{33}^s),$$

$$a_2 = -\frac{c_{13}^s r^s}{\eta_s^0} + \frac{c_{33}^s r^s}{\kappa^s c_{44}^s \eta_{s2}^0} [c_{44}^s (r^s + 1) - c_{44}^f (r^f + 1)],$$

$$a_3 = c_{33}^s \left( r^f \frac{\eta_s^0}{\eta_{f2}^0} - \frac{c_{44}^f \eta_{f2}^0}{c_{44}^s \eta_{s2}^0} \frac{r^s}{\kappa^s} \right), \ a_4 = \frac{c_{44}^f}{c_{44}^s} \frac{\eta_{f2}^0}{\eta_{s2}^0} (c_{13}^s - r^s c_{33}^s),$$

$$a_5 = \frac{1}{(c_{33}^s \eta_s^0)^2} \left\{ \left[ \bar{c}_{33}^f c_{33}^s (r^s)^2 \frac{\eta_f^0}{\eta_s^0} + r^f [r^f (c_{13}^s - c_{33}^s r^s) - c_{13}^f] \frac{c_{33}^s \eta_s^0}{\eta_f^0} \right] c_{33}^s \eta_s^0 - \bar{c}_{33}^f c_{33}^s c_{13}^s r^s \eta_f^0 \right\},$$

$$a_6 = \frac{r^f}{\eta_f^0} (c_{33}^s r^s - c_{13}^s) - \frac{\bar{c}_{33}^f}{c_{33}^s} \frac{\eta_f^0 r^s}{(\eta_s^0)^2} (c_{33}^s r^s - c_{13}^s), \ a_7 = \bar{c}_{33}^f r^f - c_{13}^f + c_{13}^s - c_{33}^s r^s,$$

$$b_1 = r^s + 1 - \frac{c_{44}^f}{c_{44}^s} (r^f + 1), \ b_2 = (r^s + 1) \left( \frac{r^f}{\eta_{f2}^0} - \frac{c_{44}^f}{c_{44}^s} \frac{\eta_{f2}^0 r^s}{(\eta_{s2}^0)^2} \right),$$

$$b_3 = \frac{c_{44}^f}{c_{44}^s} \frac{\eta_{f2}^0 r^s}{(\eta_{s2}^0)^2} + \frac{1}{c_{44}^s \eta_{f2}^0} [c_{44}^f r^f - c_{44}^s r^f (r^s + 1)] + \frac{c_{44}^f \eta_{f2}^0 (r^s)^2}{(c_{44}^s \eta_{s2}^0)^2},$$

$$b_4 = r^f \frac{\eta_{s2}^0}{\eta_f^0} - \frac{\bar{c}_{33}^f}{c_{33}^s} \frac{\eta_f^0 \eta_{s2}^0 r^s}{(\eta_s^0)^2}, \ b_5 = \frac{r^s}{\eta_{s2}^0} - \frac{r^s \eta_{s2}^0}{c_{33}^s (\eta_s^0)^2} (\bar{c}_{33}^f r^f - c_{13}^f + c_{13}^s - c_{33}^s r^s),$$

$$b_6 = \frac{r^s + 1}{c_{33}^s \eta_s^0} (\bar{c}_{33}^f r^f - c_{13}^f + c_{13}^s - c_{33}^s r^s).$$

In the above expressions, the subscript "$n$" indicating the order of vibration mode is neglected.